\documentclass[aps,prb,twocolumn,showpacs,superscriptaddress]{revtex4}
\usepackage{amssymb}
\usepackage{amsfonts}
\usepackage{amsmath}
\usepackage{bm}
\usepackage{textcomp}
\usepackage{color}
\usepackage{longtable}
\usepackage{wrapfig}
\usepackage{verbatim}

\usepackage{graphicx}% Include figure files
\usepackage{dcolumn}% Align table columns on decimal point

\begin{document}

\title{Element specific characterization of heterogeneous magnetism in (Ga,Fe)N films}

\author{I. A. Kowalik}
\email{ikowalik@ifpan.edu.pl}
\affiliation{Institute of Physics, Polish Academy of Sciences, al.~Lotnik\'ow 32/46, 02-668 Warsaw, Poland}

\author{A. Persson}
\affiliation{Department of Physics and Astronomy, Uppsala University, P.O. Box 516, 751 20 Uppsala, Sweden}

\author{M. \'A. Ni\~no}
\affiliation{Sincrotrone Trieste, S.S. 14, km 163.5, 34149 Basovizza, Trieste, Italy}

\author{A. Navarro-Quezada}
\affiliation{Institute of Semiconductor and Solid State Physics, Johannes Kepler University, Altenbergerstr. 69, 4040 Linz, Austria}

\author{B. Faina}
\affiliation{Institute of Semiconductor and Solid State Physics, Johannes Kepler University, Altenbergerstr. 69, 4040 Linz, Austria}

\author{A. Bonanni}
\affiliation{Institute of Semiconductor and Solid State Physics, Johannes Kepler University, Altenbergerstr. 69, 4040 Linz, Austria}

\author{T. Dietl}
\affiliation{Institute of Physics, Polish Academy of Sciences, al.~Lotnik\'ow 32/46, 02-668 Warsaw, Poland}
\affiliation{Institute of Theoretical Physics, University of Warsaw, 00-681 Warsaw, Poland}

\author{D. Arvanitis}
\affiliation{Department of Physics and Astronomy, Uppsala University, P.O. Box 516, 751 20 Uppsala, Sweden}

\date{\today}

\begin{abstract}

We employ x-ray spectroscopy to characterize the distribution and magnetism of particular alloy constituents in (Ga,Fe)N films grown by metal organic vapor phase epitaxy. Furthermore,
photoelectron microscopy gives direct evidence for the aggregation of Fe ions, leading to the formation of Fe-rich nanoregions adjacent to the samples surface.  A sizable x-ray magnetic circular dichroism (XMCD) signal at the Fe L-edges in remanence and at moderate magnetic fields at 300~K links the high temperature ferromagnetism with the Fe(3$d$) states. The XMCD response at the N K-edge highlights that the N(2$p$) states carry considerable spin polarization. We conclude that FeN$_{\delta}$  nanocrystals, with $\delta > 0.25$,
stabilize the ferromagnetic response of the films.

\end{abstract}

%75.50.Pp   Magnetic semiconductors
%75.10.Dg   Crystal-field theory and spin Hamiltonians
%75.70.Ak   Magnetic properties of monolayers and thin films
%75.30.Gw   Magnetic anisotropy
%81.05.Ea   III-V semiconductors
%78.70.Dm 	X-ray absorption spectra
 %75.75.Lf 	Electronic structure of magnetic nanoparticles
 %75.50.Lk 	Spin glasses and other random magnets
 %75.50.Tt 	Fine-particle systems; nanocrystalline materials
 %78.20.Ls 	Magneto-optical effects
 %75.50.Bb Fe and its alloys

\maketitle

The studies of the family of materials well known as dilute magnetic semiconductors (DMSs) have gained attention since the discovery that long range ferromagnetic coupling is
possible in materials, such as GaAs, doped with Mn, at technologically relevant temperatures.\cite{Ohno:1998_S} The mean-field $p-d$ Zener model predicts room temperature ferromagnetism in wide gap nitrides, oxides or even in diamond, provided that the concentrations of both transition metal (TM) ions substitutional of cations and delocalized or weakly localized holes is adequately high.\cite{Dietl:2000_S} It has been found, however, that due to a significant contribution of highly lying open $d$ orbitals to the bonding energy, the magnetic cations, instead of assuming random positions over the lattice sites, tend to aggregate during the epitaxy.
Embedded nanocrystals are formed, which either preserve the host structure (chemical decomposition) or  give rise to precipitates (crystallographic decomposition).\cite{Bonanni:2009_RCS,Sato:2010_RMP} Owing to a large concentration of the magnetic constituent, these nanocrystals exhibit high ordering and blocking temperatures, typically above room temperature and, therefore, control the magnetic properties of the system. The hybrid semiconductor/magnet nanocomposites formed in this way are expected to show a number of striking functionalities,\cite{Katayama-Yoshida:2007_PSSA,Dietl:2008_JAP} an example being the electromotive force generated by magnetizing zinc-blende MnAs nanocrystals buried in GaAs.\cite{Hai:2009_N} Since the TM $d$-levels reside in the gap, the aggregation process can be controlled by changing the valence of the TM ions by co-doping with donor or acceptor impurities.\cite{Kuroda:2007_NM,Bonanni:2008_PRL} It has been also suggested that the trapping of carriers by TM ions could shift the Fermi level towards an impurity band formed by native defects, whose Stoner-like magnetism becomes more relevant than the one associated directly with the magnetic ions.\cite{Coey:2008_JPD}

Further progress in this field is possible by means of nondestructive spin-sensitive and element-specific mapping of the material properties with nanometric lateral and in-depth spatial resolutions.
Here, by exploiting x-ray absorption spectroscopy (XAS) and related techniques we obtain information on the local chemistry and magnetism of Fe-doped GaN, whose previous structural and magnetization studies pointed to a highly non-uniform Fe distribution and complex macroscopic magnetic properties already for a concentration of the magnetic ions below 0.5\%.\cite{Przybylinska:2006_MSE,Bonanni:2008_PRL,Bonanni:2007_PRB,Rovezzi:2009_PRB,Navarro:2010_PRB}  In particular,  by x-ray photoemission electron microscopy (XPEEM), we demonstrate directly and in a non-destructive fashion the existence of Fe-rich nano-scale regions residing close to the surface of (Ga,Fe)N epitaxial films. We then reveal ferromagnetic features in x-ray magnetic circular dichroism (XMCD) at both the Fe L- and N K-edges. We show that Fe-N nanocrystals embedded in the GaN matrix stabilize the ferromagnetism of (Ga,Fe)N at 300~K.
Finally, we uncover the differences between the dominant FeN$_{\delta}$ phases obtained by nanocrystal aggregation, and the iron nitride films obtained by direct deposition, where $\delta \approx 0.25$.\cite{Hanke:2006_JAP,Takagi:2010_PRB}

We investigate (Ga,Fe)N films grown at the Kepler University Linz-Austria by metal organic vapor phase epitaxy (MOVPE) on c-plane sapphire substrates.\cite{Bonanni:2007_PRB,Navarro:2010_PRB}  We concentrate on two samples deposited at the same temperature $T_{\text{g}}$ of 950$^{\circ}$C but under different flow-rates of the Fe-precursor (FeCp$_{2}$): 100 and 300 standard cubic centimeters per minute (referred to as s100 and s300, respectively). The structural and magnetic properties of these samples have been thoroughly examined.\cite{Bonanni:2007_PRB,Navarro:2010_PRB}
XPEEM is performed at the Nanospectroscopy beam line of the Elettra Synchrotron facility in Trieste-Italy, while spectrally resolved XAS and XMCD measurements are carried out at the beam line D1011 of the MAX-lab synchrotron radiation laboratory, Lund-Sweden. The samples as introduced into the ultra high vacuum chamber show x-ray absorption specific to C and O atom impurities, which disappears after a few cycles of  \textit{in situ} soft sputtering with Ar$^{+}$ ions. After the cleaning procedure we clearly identify the N K- Ga L- and Fe L-edges. The measurements are performed in the total electron yield (TEY) mode by measuring the photocurrent of the sample. The effective mean path of the electrons is of the order of 2~nm, implying that the region down to 6~nm below the sample surface is probed. To asses the average Fe concentration, we employ the high photon energy atomic continuum of the Fe and N XAS spectra. A high magnitude of the Fe density, in the order of 5\%, is determined for the near surface region of sample (s300). For the sample s300 XAS gives a 20\% higher Fe content than for the s100 film.
In the XAS mode, XPEEM allows for chemical mapping of the near surface region by collecting secondary electrons emitted after photon absorption. When the incident photon energy corresponds to the absorption edge of an element on the surface or residing down to several
nanometers below, a significant increase is observed in the secondary electron emission from the areas where the specific element appears.
\begin{figure}[!h]
\begin{center}
 \includegraphics[scale=0.31,angle=0.0]{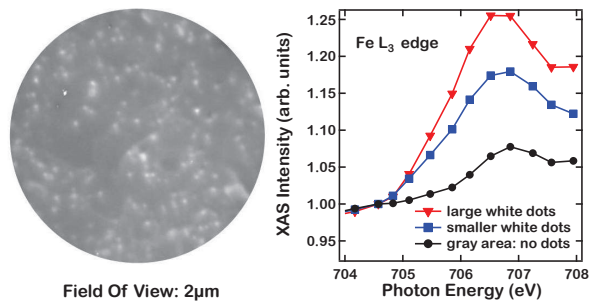}
\end{center}
\caption{(Color online) Left panel: PEEM micrograph with a 2~$\mu$m field of view for sample s300.  The photon energy is set at the Fe L$_{3}$ white line maximum. The bright spots denote areas with a high Fe concentration. The remaining regions are not entirely dark, indicating that Fe is also dissolved in the lattice. Right panel: spatially resolved XAS, corresponding to small areas within the micrograph, with differing shadowing.}
\label{PEEM_collage}
\end{figure}
Figure~\ref{PEEM_collage} shows an XPEEM micrograph for the sample s300, taken at a photon energy corresponding to the absorption maximum for the Fe L$_{3}$ edge showing the presence of iron rich nanocrystals (bright spots) with sizes of the order of 30 to 50~nm.
Though some regions 300 to 500~nm in diameter are present, where no bright spots appear, the spatially resolved XAS reported in the right panel of Fig.~\ref{PEEM_collage} confirms that these regions still contain iron. The concentration of magnetic ions is confirmed to be not homogeneous and in the regions where the distance between the Fe-rich nanocrystals is reduced, a bright halo is found connecting the structures.
These findings substantiate the overall magnetization in terms of contributions from dilute paramagnetic Fe ions and magnetic nanocrystals with high Fe-density.\cite{Navarro:2010_PRB}

\begin{figure}[!h]
\begin{center}
\includegraphics [scale=0.47,angle=0.0] {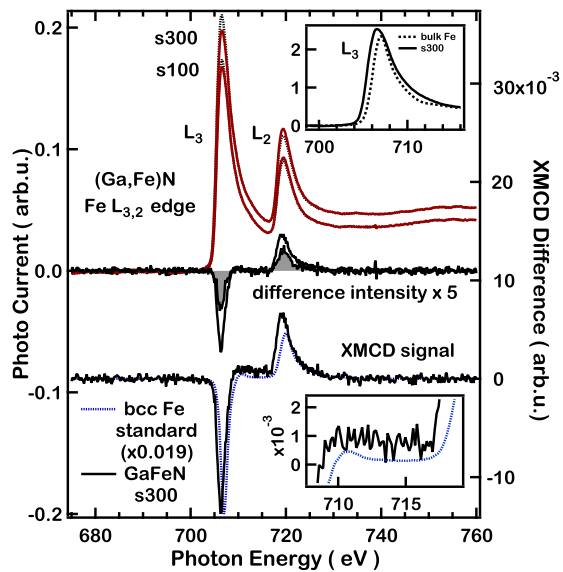}
\end{center}
\caption{(Color online) XAS and XMCD spectra in TEY mode versus photon energy. A magnetic field of $ H = \pm 800$~Oe serves to reverse the magnetization direction for the evaluation of the XMCD magnitude at 300~K.
The XAS spectra (left scale) are obtained with nearly circularly polarized x-rays. In both cases a 45$^{\circ}$ x-ray incidence angle is used. Upper inset: Fe L$_{3}$ white line together with the data for bcc Fe. The XAS spectra are shown with 1.0 subtracted. The XMCD difference spectra for (Ga,Fe)N samples s300 and s100 are reported in the lower part (left scale). As reference the spectrum of bulk bcc Fe is shown (right scale). Lower inset: the dichroic signal of the Fe (4$s$) states for sample s300 is stronger in comparison to bulk Fe.}
\label{XMCD_Fe}
\end{figure}

For averaged XAS and XMCD linear or close to circular polarization (with a degree of circular polarization of 0.8) is used. The Fe L-edges shown in Fig.~\ref{XMCD_Fe} are dominated by dipole transitions to the Fe $3d$ final states, seen in the form of intense ``white line" peaks,
each followed by a metallic continuum of final states. In the inter-peak region, the Fe $4s$ final states are also probed.

The exact photon energy of the Fe L-edges, its broadening and intensity are good markers of the electronic configuration of the Fe atoms. Here, the absolute photon energy and intensity is calibrated by using a bcc Fe reference sample measured \textit{in situ}, within 0.1 eV. The increased line width in comparison to bcc Fe - as reported in the inset to Fig.~\ref{XMCD_Fe} - is indicative of inhomogeneous broadening, consistent with Fe atoms in several different local environments and crystallographic phases, and supporting the transmission electron microscopy and synchrotron x-ray diffraction data.\cite{Navarro:2010_PRB} Interestingly, in a similar manner, by means of  XAS and XMCD, two distinct local environments for Mn atoms in (Ga,Mn)N films have been proposed, leading to a broadening of the Mn L-edge white lines when increasing the dilution of Mn in GaN.\cite{Freeman:2007_PRB} The broadening in the L edge ``white lines" of Fe observed here, corresponds to a variation in the Fe binding energy. Such an effect has been observed earlier for FeN$_{\delta}$ films.\cite{Navio:2008_PRB} The magnetic phase with $\delta = 0.25$ is linked to lower Fe binding energies, as we find here. \cite{Navio:2008_PRB} From the ``XMCD-sum rules",\cite{Carra:1993_PRL} we determine the number of $3d$ holes for (Ga,Fe)N, obtaining values larger than those found in literature for FeN$_{\delta}$, $\delta \leq 0.25$, \cite{Takagi:2010_PRB} implying a substantial degree of Fe nitridation.
The increased average number of $3d$ holes demonstrates that some of the Fe atoms are in the 3$d^{5}$ state.

\begin{table}[!h]
   \begin{center}
        \begin{tabular}{l|c|cc|cc}
        \hline   \hline
        sample   &  {number of} & \multicolumn{4}{c}{magnetic moments}                                 \\
              & {3d holes} & \multicolumn{4}{c}{($\mu_{\text{B}}$/Fe atom)}   \\
                 &   {}& \multicolumn{2}{c|}{under field} & \multicolumn{2}{c} {in remanence}  \\
                                &   $n_{h}$ & $m_{s}$  &  $m_{l}/m_{s}$       &  $m_{s}$    &  $m_{l}/m_{s}$        \\
        \hline
        (Ga,Fe)N s300           &  4.1(1) & 0.44(4)     &   0.00(1)     &  0.15(2) &   0.13(1)       \\
        (Ga,Fe)N s100           &  4.5(1) & 0.31(3)     &   0.00(1)     &          &                 \\
        \hline
        Fe bulk (bcc)           & 3.8 &             &               & 1.98     &   0.045          \\
        FeN$_{0.25}$ (2~ML)\cite{Takagi:2010_PRB}       &  3.80(15)             &  1.38(10)          &  0.13(5)    &           &   \\
        FeN$_{0.25}$ (4~ML)\cite{Takagi:2010_PRB}       &  3.80(15)             &  1.98(10)          &  0.12(5)    &           &   \\
        \hline \hline
        \end{tabular}

    \end{center}
\caption{
Parameters determined by the XMCD sum rules. The errors shown are relative, $versus$ bcc Fe.
The data under the applied field are taken at 20$^{\circ}$, 45$^{\circ}$ and 90$^{\circ}$ x-ray incidence with respect to the film plane, in remanence at $45^{\circ}$ and $20^{\circ}$. No XMCD is found in remanence at 90$^{\circ}$. The XMCD data set points to two distinct contributions to the XMCD signal.
The ratio of the orbital to spin moment in remanence is similar to the values found for two and four monolayers of FeN$_{\delta}$, $\delta \approx 0.25$,\cite{Takagi:2010_PRB} and is shown in the two lowest rows.}
    \label{magn_param}
\end{table}

The XMCD response has been recorded at 300~K under an applied in-plane field of $\pm 800$~Oe or in remanence at an x-ray incidence of 20$^{\circ}$, 45$^{\circ}$ and 90$^{\circ}$ with respect to the sample surface. The XMCD signal at various angles allows to identify the presence of different types of dichroic response, consistent with the observed broadening of the white line and the XPEEM data. Furthermore, the signal under remanence indicates an in plane easy direction.
We note that XMCD dichroic spectra, as shown in Fig.~\ref{XMCD_Fe}, in addition to the contribution from Fe($3d$) spin polarization, may contain a component originating from transitions to Fe($4s$) states, becoming evident in the spectral region between the L$_{3}$ and the L$_{2}$ lines. Actually, the so-called ``s-state" dichroic contribution is stronger for (Ga,Fe)N than for the bulk Fe, as reported in the inset to Fig.~\ref{XMCD_Fe}. This addresses the significance of the $4s-3d$ hybridization in ferromagnetic Fe atoms with N atoms as first neighbors, in particular within the
FeN$_{\delta}$ nanocrystals.
The results of the XMCD quantitative analysis are summarized in Table~\ref{magn_param}, together with XMCD data for FeN$_{0.25}$.\cite{Takagi:2010_PRB} The relatively small magnitude of the determined spin moment per Fe ion, namely $m_s \leq 0.44\,\mu_{\text{B}}$ appears inconsistent with the majority of Fe atoms in a high spin state.
The small moment value may indicate that the magnetic anisotropy of the larger FeN$_{\delta}$ nanocrystals is strong enough for those to be fully polarized in 800~Oe at 300~K. A small magnetic moment value is also consistent with the existence of smaller particles and antiferromagnetic interactions appearing at $\delta \geq 0.5$.\cite{Navarro:2010_PRB}
In order to determine the relative magnitude of the orbital and spin momentum, $m_l/m_s$, the $4s$ dichroic contribution has to be subtracted from the spectra. As reported in Tab.~I, within the experimental accuracy no orbital moment can be detected under applied field.
In contrast, in the case of the much weaker signal at remanence, an enhanced magnitude of $m_l/m_s$ is observed, similar to the one reported for FeN$_{0.25}$. This finding demonstrates that the robust ferromagnetism at 300~K originates primarily from  nanocrystals with a lower degree of nitridation. Given the correlation between the orbital moment and the magnetic anisotropy,\cite{Bruno:1989_PRB} a non zero orbital in remanence, indicates that here a minority of nanocrystals with $\delta > 0.25$ possessing a larger orbital moment is essential in stabilizing the ferromagnetic response at 300~K.

\begin{figure}[!h]
\begin{center}
 \includegraphics[scale=0.38,angle=0.0]{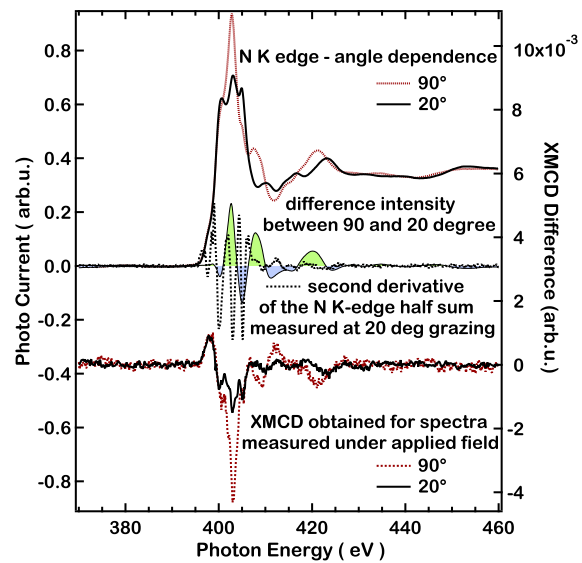}
\end{center}
\caption{(Color online) N K-edge for the (Ga,Fe)N sample s300 (left scale), with 1.0 subtracted, taken with linear x-rays at normal (90$^{\circ}$) and grazing (20$^{\circ}$) x-ray incidence $versus$ the film plane. The corresponding difference spectra are shown: negative peaks (left scale) indicate $s$ $\rightarrow$ $p_{z}$ states, positive peaks - $s$ $\rightarrow$ $p_{x},p_{y}$.  Also shown is the XMCD difference, taken with circular
%polarized
x-rays,
%obtained
for spectra
%recorded for
at
normal and grazing x-ray incidence (right scale) under an applied field of $\pm 800$~Oe.
The second derivative of the N XAS spectrum at grazing incidence is shown, as it reproduces the oscillatory part of the XMCD signal,
pointing to some itinerant character of the final states.
}
\label{XMCD_N}
\end{figure}

The N($2p$) final states are probed $via$ XAS at the N K-edge\cite{Katsikini:1999_JES} and the spectra are shown in Fig.~\ref{XMCD_N}. A shoulder at the low energy side of the edge is attributed to hybridization with the Fe electronic states. The high structural quality of the semiconductor matrix is confirmed by the dependence of the spectra versus x-ray incidence angle when linearly polarized x-rays are used. In general, K-edges lead to rather small XMCD effects.
Nevertheless, we observe a clear XMCD response from the N atoms in the magnetic field. Previously, XMCD at the N K-edge was observed for GdN.\cite{Antonov:2007_PRB} Using resonant reflectivity, N K-edge x-ray circular dichroism was detected for Fe-N in its $\gamma'$-FeN$_{0.25}$ phase.\cite{Hanke:2006_JAP}
The observed XMCD proves directly the presence of electron and spin-polarization transfer from Fe to neighboring anions, presumably $via$ spin-dependent hybridizations between N($2p$), Fe($3d$), and Fe($4s$) states.
In particular, in the presence of Fe, a clear XMCD effect is seen around 397~eV at the pre edge region of the N K-edge, which corresponds to the forbidden energy range within the band gap. Furthermore, a strong angular dependence is observed in the N K edge XMCD signal at higher energies, around 403~eV. Here, more dichroic intensity is obtained at normal x-ray incidence.
As a comparison, the GdN XMCD data indicate also a signal of a comparable intensity, about 0.5\% of the N atomic high photon energy continuum step.\cite{Antonov:2007_PRB} This appears surprising
as the concentration of N atoms
involved here in the magnetic response is expected to be much smaller, even though we probe the surface region where the density of FeN$_{\delta}$ nanocrystals is high. However, the hybridization between N($2p$) and Fe($3d$) states is expected to be  stronger than with the highly localized Gd($4f$) orbitals $via$ Gd(5$d$) states.
The XMCD signal in Fig.~\ref{XMCD_N} appears to consist of two contributions: an oscillatory part is superimposed to broader N($2p$) localized features, reemphasizing the presence of different magnetic phases. Following an earlier work\cite{Brouder:1996_PRB} we plot the second derivative of the XAS spectrum. Agreement is observed  with the oscillatory part of the XMCD signal, suggesting that the probed final states possess a considerable degree of itinerant character. One determines from the spectra of Fig.~\ref{XMCD_N} an exchange splitting energy of about 0.9~eV that falls within the expected range for itinerant magnetism, eventually indicating that the FeN$_{\delta}$ nanocrystals polarize also atoms in their near vicinity, giving reason for the large XMCD signal and its angle dependence.

In summary, we characterize the distribution, chemistry and magnetism of FeN$_{\delta}$ nanocrystals  aggregating in a self-organized manner in epitaxial layers of (Ga,Fe)N. These nanocrystals have a diameter below 50~nm, reside primarily close to the surface and are characterized by various degrees of nitridation. We quantify the impact of the degree of nitridation on the orbital moment and the stability of ferromagnetism at 300~K. Depending on the growth conditions, (Ga,Fe)N  may serve to develop nanocomposite spintronic systems characterized by either robust room temperature ferromagnetism  ($\delta \leq 1/3$) or strong antiferromagnetic interactions ($\delta > 1/2$).

We acknowledge the EC support through the FunDMS Advanced Grant of the ERC within the ``Ideas" 7th Framework Programme, the Austrian Fonds zur F\"{o}rderung der wissenschaftlichen Forschung-FWF (P22477, P18942, P20065 and N107-NAN), the Swedish Research Council, the EC ARI program for access to MAX-lab, A. Locatelli and T. Onur Mentes from Elettra for support in the PEEM work, M. Sawicki for communicating the  SQUID results and for numerous discussions.

%\bibliography{References_Kowalik}
%\bibliographystyle{apsrev}

\end{document}